\documentclass{ws-p8-50x6-00}
\usepackage{graphicx}
\usepackage{amsmath,amssymb}
\newcommand{\sumk}{\sum_{{\mathbf k}}}
\newcommand{\phik}{\varphi_{\mathbf{k}}}
\newcommand{\ek}{\epsilon_{\mathbf{k}}}
\newcommand{\ekq}{\epsilon_{\mathbf{k-q}}}
\newcommand{\phikq}{\varphi_{{\mathbf{k}}-{\mathbf{q}}/2}}
\newcommand{\Ek}{E_{\mathbf{k}}}

\newcommand{\uk}{u_{\mathbf{k}}}
\newcommand{\vk}{v_{\mathbf{k}}}

\begin{document}

\title{Magnetic field effects on $T_c$ and the pseudogap onset
  temperature in cuprate superconductors}

\author{Qijin Chen}

\address{National High Magnetic Field Laboratory, 1800 East Paul Dirac
  Drive, Tallahassee, Florida 32310}

\author{Ying-Jer Kao, Andrew P. Iyengar, and K. Levin}

\address{James Franck Institute, University of Chicago, Chicago,
  Illinois 60637}
\maketitle

\abstracts{
  We study the sensitivity of $T_c$ and the pseudogap onset temperature,
  $T^*$, to low fields, $H$, for cuprate superconductors, using a
  BCS-based approach extended to arbitrary coupling.  We find that $T^*$
  and $T_c$, which are of the same superconducting origin, have very
  different $H$ dependences. The small coherence length makes $T^*$
  rather insensitive to the field.  However, the presence of the
  pseudogap at $T_c$ makes $T_c$  more sensitive to $H$. Our
  results for the coherence length $\xi$ fit well with existing
  experiments. We predict that very near the insulator $\xi$ will
  rapidly increase.
}


The pseudogap phenomena have been a great challenge to condensed matter
physicists since last century. Yet there has been no consensus on the
origin of the pseudogap and its relation to the superconducting order
parameter.  Theories about the pseudogap physics fall into two
categories: (1) precursor versus (2) non-precursor superconductivity.
For the former, pseudogap forms as a consequence of precursor
superconducting pairing, and therefore, shares the same origin as the
order parameter. In contrast, for the latter category, pseudogap has a
different origin, e.g., a hidden DDW (d-density wave) order.\cite{DDW}

On the other hand, experiment has revealed different behaviors of $T_c$
and the pseudogap onset temperature $T^*$ in magnetic
fields.\cite{Experiment} However, there is still no proper theoretical
explanation. This difference has been used as evidence against precursor
superconductivity scenarios. Here we show that it can be well explained
within the present precursor superconductivity
theory.\cite{Kosztin,PRL98,PRL00}

Our calculation is based on an extension of BCS theory which
incorporates incoherent pair excitations. These pair excitations become
increasingly important at large coupling $g$, and lead to a pseudogap in
the single-particle excitation spectrum, as seen in the cuprates. As the
temperature increases from $T=0$, these pairs can survive a higher
temperature ($>T_c$) than the condensate, until they are completely
destroyed by the thermal effect at $T^*$. In agreement with experiment,
we find that $T^*$ and $T_c$ have very different field ($H$)
dependences. The small coherence length ($\xi$) makes $T^*$ rather
insensitive to the field.  However, the presence of the pseudogap at
$T_c$ (at optimal and under- doping) makes $T_c$ relatively more
sensitive to $H$.  Our results for the coherence length $\xi$ fit well
with existing experiments.  Furthermore, we predict that very near the
insulator $\xi$ will rapidly increase.

We first consider the zero magnetic field case. We include, in addition
to time-reversal state pairing, finite center-of-mass momentum pair
excitations in the problem and then treat the interrelated single- and
 two-particle propagators self-consistently. We truncate the infinite
series of equation of motion at the three-particle level, and then
factorize the three-particle Green's function $G_3$ into single-particle
($G$) and two-particle ($G_2$) Green's functions.\cite{Kadanoff}

Here we consider an electron system near half filling
on a quasi-two dimensional (2D) square lattice, with tight-binding
dispersion $\ek$.  The electrons interact via a separable potential
$V_{\mathbf{k,k^\prime}}=g\varphi_{\mathbf{k}}\varphi_{\mathbf{k^\prime}},$
where $\varphi_{\mathbf{k}}= \cos k_x-\cos k_y$ (for $d$-wave). We use a
T-matrix approximation for the self-energy, and have
\begin{equation}
\Sigma(K) = G_0^{-1}(K)-G^{-1}(K) 
  = \sum_Q t(Q)\,G_0(Q-K)\,\varphi^2_{{\bf k}-{\bf q}/2} \;,
\end{equation}
\begin{equation}
t(Q) = -\frac{|\Delta_{sc}|^2}{T}\delta(Q)\theta(T_c-T) +
\frac{g}{1+g\chi(Q)} \;,
\end{equation}
where $\Delta_{sc}$ is the order parameter, and
$\chi(Q)=\sum_{K}G(K)G_0(Q-K)\,\varphi^2_{{\bf k}-{\bf q}/2} $.
For small $Q\ne 0$, $t(Q)$ can be written in a standard propagator form.
The pseudogap is given by $\Delta_{pg}^2\equiv -\sum_Q t(Q)$, the total
gap\cite{PRL98} by $\Delta = \sqrt{\Delta_{sc}^2 + \Delta_{pg}^2}\:$, and
the quasiparticle dispersion by $E_{\bf k} = \sqrt{\epsilon_{\bf k}^2 +
  \Delta^2 \varphi_{\bf k}^2}\:$.

$T_c$ is determined by the superconducting instability condition
$1+g\chi(0)=0$, in conjunction with the number constraint $n=2\sum_K
G(K)$. We obtain a set of three equations.\cite{PRL98} Taking into
account that the cuprates is close to the Mott insulator and thus
in-plane hopping $t_\parallel (x)=t_0 (1-n)=t_0x$, where $x$ is the
doping concentration, we solve for $T_c$, $\Delta$, chemical potential
$\mu$, and $\Delta_{pg}$.  The results for $T_c$, $\Delta_{pg}(T_c)$,
and $\Delta(T=0)$ as a function of $x$ are summarized in Fig.~1(a). Our
predictions fit experiment well with $-g/4t_0=0.045$ and $t_0\approx
0.6$ eV. For more details, see Refs.~3--5.

\begin{figure}[t]
\vspace*{1ex}
\centerline{\includegraphics[bb=150 129 392 282, width=2.3in]{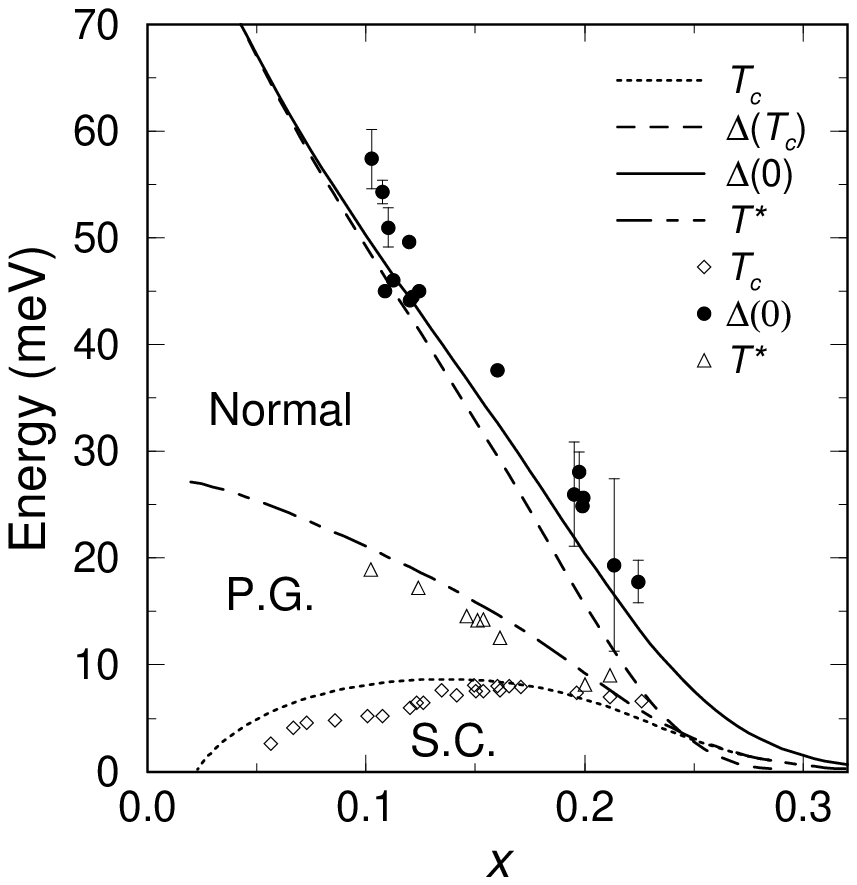} 
\hfill
\includegraphics[bb=0 20 634 679, width=2.3in, clip]{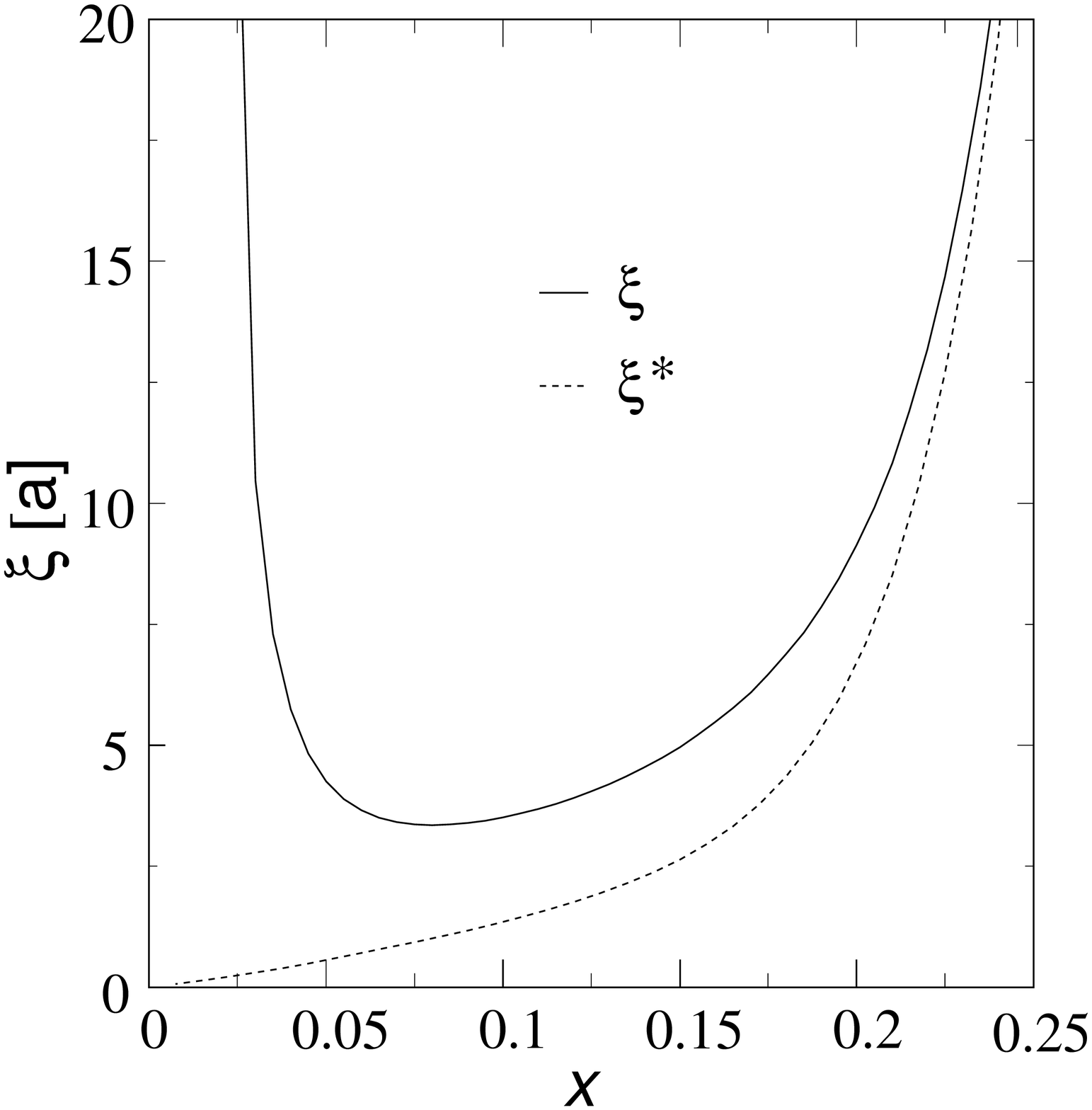}}
\vskip -1.2in
\hspace*{1.9in} \textbf{(a)} \hspace*{1.1in} \textbf{(b)}
\vskip 0.9in
\caption{(a) Calculated cuprate phase diagram. $T^*$ was estimated using
  the BCS mean-field solution.  Experimental data are taken from:
  ($\bullet$) Ref.~7; ($\diamond$) Ref.~8; ({\scriptsize $\triangle$})
  Ref.~9.  For more details see Ref.~4.  Note here $\Delta(0)$ has been
  multiplied by 2 to compare with experiment, since
  $\varphi_{\mathbf{k}}=2$ at $\textbf{k}=(\pi, 0)$.  (b) Magnetic length
  scales associated with $T_c$ and $T^*$ as a function of doping
  concentration in the cuprates. }
\end{figure}

In a finite field, the Ginzburg-Landau free energy functional near $T_c$
can be expanded to quadratic order in the order parameter $\Delta_{sc}$:
\begin{equation}
  F\sim \left(\tau_0(T)+\eta^2 \left(\frac{\nabla}{i}-\frac{2e{\mathbf
          A}}{c}\right)^2\right) \left|\Delta_{sc}\right|^2 ,
\end{equation}
where
$\displaystyle{\tau_0(T)=\bar{\tau}_0\left(1-\frac{T}{T_c}\right)}$, and
$\displaystyle{ -
  \left.\frac{1}{T_c}\frac{dT_c}{dH}\right|_{H=0}=\frac{2\pi}{\Phi_0}
  \xi^2= \frac{2\pi}{\Phi_0}\frac{\eta^2}{\bar{\tau}_0} }$. As an
estimate, one has $H_{c2}(0) \approx \Phi_0/(2\pi\xi^2) $. ($\Phi_0=hc/2e$ is
the flux quantum).

For 3D weak coupling (BCS), $\bar{\tau}_0=N(0)$ and the phase stiffness
$\eta^2=N(0) 7 \zeta(3)/48 \pi^2 (v_F/T_c)^2$.  Therefore $N(0)$ is
canceled in $\xi_{BCS}^2=7 \zeta(3)/48\pi^2({v_F}/{T_c})^2$.  In
general, $\tau_0$ and $\eta^2$ can be determined from the expansion of
$t^{-1}(Q)$:
\begin{equation}
  \tau_0=\frac{1}{g}+\chi({\mathbf 0},0), \qquad \eta^2=\frac{1}{2}
  \left. \sqrt{ {\rm det} \left[ \partial_{q_i} \partial_{q_j} \chi(Q)
      \right]} \, \right|_{Q=0} .
\end{equation}

In weak field, $T \gg eH/mc$, we use semiclassical phase approximation
to treat the single-particle and pair propagators.  Both the
single-particle and pair momenta can be modified by the interaction with
the field. But formally, the Dyson's equation remains the same, and the
superconducting transition is still determined by the pairing
instability (Thouless) condition:
$ g^{-1} + \hat{\chi}(0) =0 \approx
\tau_0 + \eta^2 \cdot \left(\frac{2e}{c} H\right)
$, 
 where
$\hat{\chi}(Q)$ is the pair susceptibility in the field.  To linear
order in $H$, we can evaluate $\tau_0$ and $\eta^2$ at $H=0$.

At $T^*$, the pseudogap is zero, only the bare Green's function is
involved, $\chi_0(Q)=\sum_{K}G_0(K)G_0(Q-K)\,\varphi^2_{{\bf k}-{\bf
    q}/2}$. We have
\begin{equation}
 - \left.\frac{1}{T^*} \frac{dT^*}{dH}\right|_{H=0}
  =\frac{\eta^{*2}}{\bar{\tau}_0^{*}}\frac{2\pi}{\Phi_0}\;, \quad
  \xi^{*2} =\frac{\eta^{*2}}{\bar{\tau}_0^*} \;,
\end{equation}
where $\displaystyle{ \bar{\tau}_0^* = \sumk \phik^2 \left[ -f'(\ek) +
    \frac{d\mu}{dT} \frac{T}{\ek} \left(\frac{1-2f(\ek)}{2\ek} +
      f'(\ek)\right) \right]_{T=T^*} }$

At weak coupling (for $s$-wave $\varphi_{k_F}=1$), we recover the BCS
limit: $\mu \approx E_F $, $\bar{\tau}_0^*=-\sumk f'(\ek)\phik^2\approx
N(0)\varphi_{k_F}^2\approx N(0)$.  And ${\eta^*}^2 $ is determined by
expanding $\displaystyle{ \chi_0({\bf q},0)=\sumk
  \frac{1-f(\ek)-f(\ekq)}{\ek+\ekq} \phikq^2 }$ to the $q^2$ order.

At large $g$ (for the underdoped cuprates), $\Delta_{pg}(T_c)$ is large.
Noticing that $T^*$ is very weakly $H$ dependent in the strong
pseudogap regime [see Fig.~1(b)], and that $T^* \propto \Delta_{pg} $ in
zero field, we assume $\Delta_{pg} $ is only weakly $H$ dependent. In
other words, only the superconducting order parameter is strongly
coupled to the field. Then we obtain $ \bar{\tau}_0\approx -\sumk
\phik^2 f'(\Ek)$, which decreases rapidly as $\Delta_{pg}$ increases.
$\eta^2$ is obtained by expanding to the $q^2$ order  
\begin{equation} 
\chi({\bf q},0)=\sumk \left[\frac{1-f(\Ek)-f(\ekq)}{\Ek+\ekq}\uk^2
    -\frac{f(\Ek)-f(\ekq)}{\Ek-\ekq} \vk^2 \right]\phikq^2.  
\end{equation}

At weak coupling, $\xi$ and $\xi_{BCS}$ coincide. But they split apart
as $g$ increases and the pseudogap opens (see Ref.~10 
for details). In Fig.~1(b), we plot the doping dependence of the
calculated $\xi$ and $\xi^*$. At large $x$ (overdoping, weak coupling),
the two are nearly equal. But for underdoping, while $\xi^*$ continues
to decrease with decreasing $x$, $\xi$ remains nearly flat for a broad
range of $x$ until its final rapid increase toward the insulator limit.
Since $dT/dH \propto \xi^2$, $T^*$ is rather insensitive and $T_c$ is
relatively more sensitive to $H$.  These results are in agreement with
experimental observations.\cite{Experiment,Exp2}

This work was supported by NSF-MRSEC, grant No.~DMR-9808595, and by the
State of Florida (Q.C.).

\end{document}